# Observational constraints on thawing quintessence scalar field model

Fereshteh Felegary*

*Faculty of Physics, Shahrood University of Technology, P.O. Box 3619995161 Shahrood, Iran.*

Kazuharu Bamba†

*Faculty of Symbiotic Systems Science, Fukushima University - Fukushima 960-1296 - Japan.*

(Dated: March 1, 2024)

Thawing quintessence scalar field models with the various potential forms to explain the late-time cosmic acceleration are compared to the $\Lambda$CDM model in detail by analyzing cosmological parameters with a set of observational data including $H(z)$, BAO, CMB, SNIa, BBN, and $f(z)\sigma_8$ at the background and the perturbation levels. At low redshifts for the thawing quintessence scalar field models, the growth rate of the cosmic structure is significant. By utilizing a standard Markov Chain Monte Carlo (MCMC) procedure based on the recent expansion and the growth observational data with the statistical values of the Akaike and the Bayesian information criteria, we discuss the consistency of the thawing quintessence scalar field models with the set of different potentials with the observational data. The main consequence of this work is that despite the various considered potential forms that are very popular in the literature, we should be looking for consistent potential forms with observational data.

## I. INTRODUCTION

In recent years, one of the most remarkable discoveries in modern cosmology is that our universe is experiencing an accelerated expansion phase. A wide range of set of observational data including Baryonic Acoustic Oscillation (BAO) [1–4], Cosmic Microwave Background radiation (CMB) [5, 6], Large Scale Structure (LSS) [7–9], and SuperNovae type Ia (SNIa) [10–12] affirm the current accelerated expansion of the universe. This cosmic accelerated expansion of our universe can be due to the presence of a perfect fluid with negative pressure called dark energy (DE) or due to the modification of gravity itself. The simplest model supposes that the dark energy is connected with the vacuum energy which is called cosmological constant ($\Lambda$) with the constant equation of state parameter (EoS) described as $\omega_\Lambda = p_\Lambda/\rho_\Lambda = -1$. This model suffers from two main problems: one is the fine-tuning problem and another one is the cosmic coincidence problem [13–25]. Thus, various models for dark energy have been proposed in which their equation of state parameters change with time. Some of the most significant category of these models are Tachyon [26, 27], Quintessence [28, 29], Phantom [30–32], Quintom [33, 34], K-essence [35, 36], Agegraphic [37], New agegraphic [38, 39], and Holographic [40–42] models. As another option, the accelerated expansion of the Universe could be explained in the framework modified theories of gravity [21, 43, 44].

Discovering of the accelerated expansion of the universe [10, 11], the effect of dark energy in cosmic history has been one of the most significant challenges in cosmology in the recent decades. Despite many attempts from both the observational and the theoretical features, the nature of dark energy is still unknown. So many various models for dark energy with a varying equation of state parameter have been suggested. Therefore, it is evident that more general classes of models allow time evolution of dark energy like scalar field models. Scalar fields arise from string theory and particle physics. Therefore, these could be the appropriate candidates to describe the nature of the dark energy if they are sufficiently strongly self-intracting. Scalar field models can also reduce the fine tuning and coincidence problems and prepare a convenient alternative to cosmological constant [45]. With these interpretations, scalar field models could be categorized as two classification which depend on their potential: one is the fast-roll model and the other one is the slow-roll model. These models are called freezing model and thawing model, respectively. In the fast-roll models, the potential is steep and the scalar field tracking the background is subdominant for most of the evolution history. This field becomes dominant and drives the acceleration of the universe at late times which is known as the tracker. In the slow-roll model, the field kinetic energy is much smaller than its potential energy. Generally, this model has sufficiently flat potential like an inflation. Due to the large Hubble friction, the field is almost frozen at early times and its equation of state parameter is $-1$. Also, its energy density is almost fixed and has an unimportant contribution to the total energy density of the universe. Due to the expansion of the universe, the radiation and the matter quickly dilute and the scalar field energy density becomes comparable to the background energy density. In this model, the field breaks away from its frozen state and its equation of the state parameter changes slowly as $\omega > -1$. It is worth mentioning in order to gain a fixed late-time evolution, this model requires some degree of fine-tuning of the initial conditions [46].

---
* f.felegary@shahroodut.ac.ir
† bamba@sss.fukushima-u.ac.jp

The subject of structure formation in the universe is a significant problem in cosmology. Dynamical dark energy models such as the quintessence model, not only reduce the theoretical of the cosmological constant but also these scalar field models straightly change the dynamics of the Hubble flow. Also, in addition to the expansion of the universe, dark energy models can affect the power spectrum of matter and large scale structures through its fluctuations [8]. Generally, in dynamical dark energy models that equation of state parameters change with time, one can suppose that the dark energy perturbations act in a similar manner to matter [47–50]. The main parameters to explain perturbations on the growth of large-scale structures in the universe are the equation of state parameter of dark energy, $\omega_\phi$, and the effective sound speed, $c_{eff}^2$. Indeed, effective sound speed is defined as $c_{eff}^2 = \delta p_\phi / \delta \rho_\phi$ in order to explain the clustered and the smooth dark energy models. Originally, if dark energy is smooth then we will have $c_{eff}^2 = 1$, while we utilize $c_{eff}^2 = 0$ for clustered dark energy models. In the clustered dark energy models, the effective sound speed of dark energy is much smaller than the speed of light. Thus, the dark energy perturbations inside the Hubble radius can increase through gravitational instability such as in pressure-less matter fluctuations. In the smooth models, the effective sound speed of dark energy is close to the speed of light. So the dark energy perturbations inside the Hubble radius cannot increase through gravitational instability [48, 51–53].

The recent observations indicate that the equation of state parameter for dark energy does not significantly deviate from $\omega_\phi = -1$ around the present era [54, 55]. This kind of equation of state parameter can be acquire in dynamical models such as thawing scalar fields. Motivated by this fact, it was investigated the quintessence scalar field model with approximately flat potentials which satisfies the slow-roll conditions [56]. In according to the slow-roll conditions, It was demonstrated that a scalar field with a variety of potentials evolves in a similar procedure. It can also conclude a general interpretation for the equation of state parameter for the scalar fields. The similar results were made for the phantom and tachyon scalar field [57–59]. In according to the slow-roll conditions, It was displayed that all of these models have the same equation of state parameter and therefore can not be recognized at the background level of cosmology. The important supposition for getting at this main consequence was the fulfillment of the slow-roll conditions for the potentials. In recent years, there has been a major development in our understanding of the role of the thawing scalar field models of dark energy. Sen and his Colleagues [59] have focused on observational quantities like the Hubble parameter and the luminosity distance for the thawing tachyon and the quintessence scalar field models of dark energy. It has been investigated the different classes of scalar fields of dark energy with a variety of potential belonging to the thawing type like $V(\phi) = \phi^{-1}, \phi^{-2}$ [58]. Also, the general evolution of spherical over-densities for the thawing class of dark energy models has been studied by Devi et al [60]. As it mentioned, the thawing dark energy models are specified in such a method that the scalar field is frozen in the early universe by very large amounts of the Hubble damping because of the expansion of the universe. While the universe expands, the Hubble parameter reduces and the Hubble damping and the scalar field begin evolving slowly down its potential. Hence, at the first the equation of state parameter starts with $\omega_\phi = -1$ and in the later time, it slowly leaves from this amount. Since the subject of the structure formation in the universe is a significant problem in cosmology and the thawing scalar fields of dark energy are interesting topics in recent years, it caused a motivation for us to study the thawing quintessence scalar field model and to compare it with the recent observational data.

Generally, it is supposed that there are no density perturbations in the quintessence scalar field model on the cluster scales. The reason for this supposition is that in the linear perturbation theory, the mass of this field model is very small, so it does not feel overdensities of the size 10 Megaparsec or smaller [61, 62]. Therefore, in this work we just investigate the thawing quintessence scalar field model at the smooth perturbations. Since the scalar field models depend on their potential, in order to study these models, a lot of potentials have been introduced. In recent years, there have been some studies about the consistency of the scalar field models with observational data which importance of this issue shows [62–64]. As it is mentioned, the thawing quintessence scalar field model depends on its potential, so we will consider the set of the different potentials as $V = \phi^\alpha$ and $e^{\alpha\phi}$ which $\alpha$ is an arbitrary constant number. It is necessary to point out that $V = \phi^\alpha$ is included $V_1 = \phi, V_2 = \phi^2, V_3 = \phi^{-2}$ and $e^{\alpha\phi}$ is included $V_4 = e^\phi, V_5 = e^{-\phi}$ [64].

The structure of this paper is formed as follows: In Section 2, the thawing quintessence scalar field Model and its evolution are introduced and its evolution is compared with the evolution of the $\Lambda$CDM model. In Section 3, we investigate the growth of density perturbations in the thawing quintessence scalar field Model at smooth perturbation level and compare our consequences with the $\Lambda$CDM model. In Section 4, using the current cosmological data, we perform likelihood statistical analysis in the smooth perturbation and background levels and fit the model with the latest observational data. Finally, we sum up our results of this paper in Section 5.

## II. THAWING QUINTESSENCE SCALAR FIELD MODEL

The quintessence scalar field model is described by a minimally coupled scalar field, $\phi$. The action for the quintessence is obtained by [65]

$$S = \int d^4x \sqrt{-g} \Big[ -\frac{1}{2} g^{\mu\nu} \partial_\mu \phi \partial_\nu \phi - V(\phi) \Big], \quad (1)$$

where $V(\phi)$ is the potential of the quintessence scalar field. In a framework of the flat spacetime Friedmann-Robertson-Walker (FRW) metric, the variation of the action Eq. (1) with respect to $\phi$, the equation of motion for the quintessence scalar field is given by [65]

$$\ddot{\phi} + 3H\dot{\phi} + \frac{dV}{d\phi} = 0, \quad (2)$$

where the symbol $\cdot$ denotes the derivative with respect to the cosmic time, $t$. Also, $H$ is called the Hubble parameter and it is defined as

$$H = \frac{\dot{a}(t)}{a(t)}. \quad (3)$$

where $a(t)$ is the scale factor. The variation of the action Eq. (1) with respect to $g^{\mu\nu}$, the energy-momentum tensor of the quintessence scalar field model is given by [65]

$$T_{\mu\nu} = -\frac{2}{\sqrt{-g}} \frac{\delta S}{\delta g^{\mu\nu}}$$
$$= \partial_\mu \phi \partial_\nu \phi - g_{\mu\nu} \Big[ \frac{1}{2} g^{\alpha\beta} \partial_\alpha \phi \partial_\beta \phi + V(\phi) \Big]. \quad (4)$$

Using Eq. (4), in a framework of the flat spacetime FRW metric, one can obtain the pressure and the energy density of the scalar field [65]

$$p_\phi = T_i^i = \frac{\dot{\phi}^2}{2} - V(\phi), \quad (5)$$

$$\rho_\phi = -T_0^0 = \frac{\dot{\phi}^2}{2} + V(\phi). \quad (6)$$

Now, we consider a universe is filled with dark energy, dark matter and radiation, the first Freidmann equation is [65]

$$H^2 = \frac{1}{3M_{pl}^2} \Big( \rho_\phi + \rho_m + \rho_r \Big), \quad (7)$$

where $M_{pl} = 1/\sqrt{8\pi G}$ is the reduced Planck mass. Also, $\rho_\phi$, $\rho_m$ and $\rho_r$ are the energy density of scalar field, pressureless dark matter and radiation, respectively. We consider that there are no interactions between the cosmic fluids. Thus, the continuity equations are given by [65]

$$\dot{\rho}_\phi + 3H\rho_\phi(1 + \omega_\phi) = 0, \quad (8)$$

$$\dot{\rho}_m + 3H\rho_m = 0, \quad (9)$$

$$\dot{\rho}_r + 4H\rho_r = 0, \quad (10)$$

here $\omega_\phi$ is the equation of state (EoS) parameter of dark energy which is defined as [65]

$$\omega\phi = \frac{p_\phi}{\rho_\phi}, \quad (11)$$

The fractional energy densities are defined as [65]

$$\Omega_\phi = \frac{\rho_\phi}{3M_{pl}^2 H^2}, \quad (12)$$

$$\Omega_m = \frac{\rho_m}{3M_{pl}^2 H^2}, \quad (13)$$

$$\Omega_r = \frac{\rho_r}{3M_{pl}^2 H^2}, \quad (14)$$

Using Eqs. (7), (12), (13) and (14), the first Friedmann equation is written as

$$\Omega_\phi + \Omega_m + \Omega_r = 1. \qquad (15)$$

Now, using Eqs. (7), (12), (13), (14), $\rho_r = \rho_{r_0} a^{-4}$, $\rho_m = \rho_{m_0} a^{-3}$ and the relation between the scale factor $a$ and the redshift $z$, $a = (1+z)^{-1}$, the dimensionless Hubble parameter is given by

$$E^2(z) = \frac{H^2(z)}{H_0^2(z=0)} = \frac{\Omega_{r_0}(1+z)^4 + \Omega_{m_0}(1+z)^3}{1 - \Omega_{de}(z)}, \qquad (16)$$

where $H_0$ is the Hubble parameter at the present time and $\Omega_{m_0}$ and $\Omega_{r_0}$ are the present amounts of dimensionless densities for dark matter and radiation, respectively.

Using Eqs. (5), (6) and (11) for the quintessence scalar field model, the EoS parameter is described by

$$\omega_\phi = \frac{\dot\phi^2 - 2V}{\dot\phi^2 + 2V}. \qquad (17)$$

We will pursue a similar method which is followed in [66, 67] and use the variables $\lambda$, $\Gamma$, and $\gamma$ as follows [66, 67]

$$\lambda = -\frac{1}{V}\frac{dV}{d\phi}, \qquad (18)$$

$$\Gamma = V\frac{\frac{d^2V}{d\phi^2}}{(\frac{dV}{d\phi})^2}, \qquad (19)$$

$$\gamma = 1 + \omega_\phi. \qquad (20)$$

Now, taking the scale factor derivative of Eq. (18) and using Eqs. (12), (18), (19), and (20), we can obtain [66, 67]

$$\lambda' = -\sqrt{3}\lambda^2(\Gamma - 1)\sqrt{\gamma\Omega_\phi}, \qquad (21)$$

Taking the scale factor derivative of Eq. (20) and using Eqs. (12), (18), and (20), we can obtain [66, 67]

$$\gamma' = -3\gamma(2-\gamma) + \lambda(2-\gamma)\sqrt{3\gamma\Omega_\phi} \qquad (22)$$

Also, taking the scale factor derivative of Eq. (12) and using Eqs. (7), (8), (9), (12), and (20), one can obtain [66, 67]

$$\Omega'_\phi = \Omega_\phi[3(1-\gamma)(1-\Omega_\phi) + \Omega_r]. \qquad (23)$$

where prime denotes the derivative with respect to $\ln a$. For the quintessence scalar field model, the evolution Eqs. (21), (22), and (23) are an autonomous system of equations involving the observable parameter $\Omega_\phi$ and $\gamma$. By giving the initial conditions for the parameters $\Omega_\phi$, $\gamma$, and $\lambda$, we can solve this system of equations numerically for the various potentials. As it is mentioned, we are interested in studying the thawing model. In the thawing model, the equation of state parameter is initially frozen at $\omega_\phi = -1$. Therefore, for our aim, $\gamma$ is required to be 0. Moreover, we suppose that the slow-roll conditions are strongly broken for the scalar field potentials i.e. $\lambda_{initial} \sim 1$. We should point out that the slow-roll conditions are satisfied for models i.e. $\lambda \ll 1$ [66, 67]. Generally, at early times, the contribution of the scalar field energy density of the universe is negligible, however, one has to fine-tune the initial amount of $\Omega_\phi$ to have its accurate contribution at present. With the above initial conditions, we can solve the autonomous system of equations for the thawing quintessence scalar field model. For solving the autonomous system of equations, we consider the set of the different potentials [64]

$$V = \phi^\alpha, \quad e^{\alpha\phi} \implies V_1 = \phi, \quad V_2 = \phi^2,$$
$$V_3 = \phi^{-2}, \quad V_4 = e^\phi,$$
$$V_5 = e^{-\phi}, \qquad (24)$$

where $\alpha$ is an arbitrary constant number. Using Eqs. (19) and (24), we can obtain the values of $\Gamma$, respectively [64]

$$V = \phi^\alpha \implies \Gamma_i = 1 - \frac{1}{\alpha}$$
$$\implies \Gamma_1 = 0, \quad \Gamma_2 = \frac{1}{2}, \quad \Gamma_3 = \frac{3}{2},$$
$$V = e^{\alpha\phi} \implies \Gamma_i = 1$$
$$\implies \Gamma_4 = 1, \quad \Gamma_5 = 1$$
$$\implies \Gamma_{4,5} = 1. \qquad (25)$$

At the present era, we also consider $\Omega_{\phi_0} = 0.7$ for all chosen values of $\Gamma$, Eq. (25). Now, using Eqs. (16), (20), (24), (25) and numerical solution of Eqs. (21), (22) and (23), we can plot the evolution of $\omega_\phi(z)$, $\Omega_\phi(z)$ and $E(z)$ as the function of the cosmic redshift for the thawing quintessence model with the set of the different potentials in Figure (1). We then compare the evolution of them with the standard $\Lambda CDM$ model.

Figure (1-a) shows that the EoS parameter of the thawing quintessence model with the set of the different potentials has different behavior when their evolution approaches the present time and all models approach $-1$ at high redshift. Also, Figures (1-b) and (1-c) indicate that the evolution of $\Omega_\phi(z)$ and $E(z)$ of the thawing quintessence model with the set of the

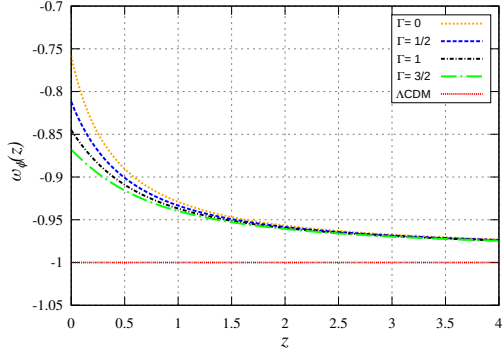

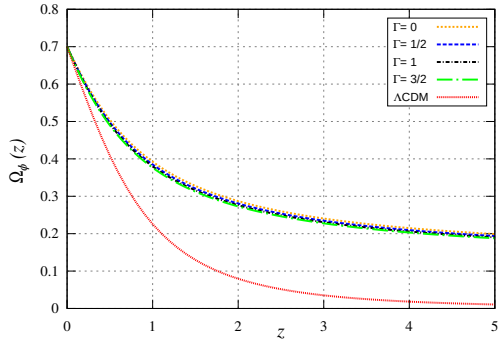

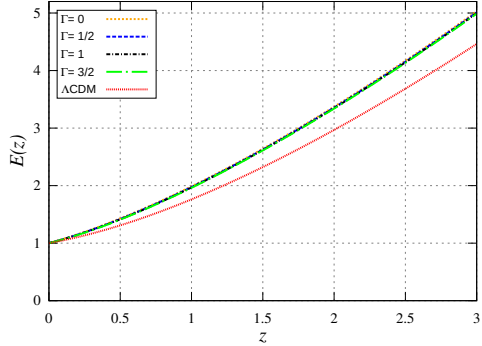

FIG. 1: Evolution of (a) $\omega_\phi(z)$, (b) $\Omega_\phi(z)$ and (c) $E(z)$ as the function of the cosmic redshift for the thawing quintessence model of dark energy with the set of the different potentials.

different potentials have almost the same behavior in all times but their evolution behaves differently from the standard $\Lambda$CDM model.

For a universe is filled with dark energy, dark matter, and radiation, the deceleration parameter is de-

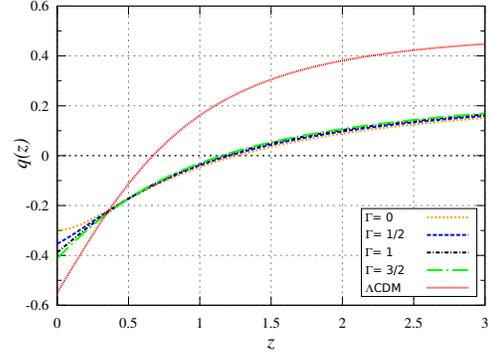

FIG. 2: Evolution of $q(z)$ as the function of the cosmic redshift for the thawing quintessence model of dark energy with the set of the different potentials.

scribed

$$q(z) = -1 - \frac{\dot{H}}{H^2} = \frac{3}{2}\Omega_\phi\omega_\phi + \frac{1}{2}(1+\Omega_r). \quad (26)$$

In Eq. (26), if $q(z) < 0$, then we will be able to experience a universe with accelerating expansion and if $q(z) > 0$, then we will be able to experience a universe with decelerating expansion [10, 11]. Using Eqs. (20), (24), (25) and numerical solution of Eqs. (21), (22), (23) and replacing in Eq. (26), we can plot the evolution of $q(z)$ as the function of the cosmic redshift for the thawing quintessence model with the set of the different potentials in Figure (2). We then compare the evolution of them with the standard $\Lambda$CDM model. For plotting this Figure, we assume $\Omega_{\phi_0} \approx 0.7$ and $\Omega_{r_0} \approx 9 \times 10^{-5}$. Figure (2) shows that the evolution of the deceleration parameter of the thawing quintessence model with the set of the different potentials has almost the same behavior at all times but their evolution behaves differently from the standard $\Lambda$CDM model. In this Figure, we can see that the evolution of the deceleration parameter of the thawing quintessence model with the set of the different potentials enters to accelerating phase earlier than the standard $\Lambda$CDM model. This is the reason that in these models, the evolution of the Hubble parameter is faster than the evolution of the Hubble parameter in the standard $\Lambda$CDM model.

Now, using a set of the observational data consisting of the Hubble data [68–78], we can plot the evolution of the Hubble parameter ($H(z)$) as a function of cosmic redshift for the thawing quintessence model with the set of the different potentials. We then compare the evolution of them with the standard $\Lambda$CDM model. Figure (3) shows that the evolution of $H(z)$ of the thawing quintessence model has the same behavior as the standard $\Lambda$CDM model at low redshifts.

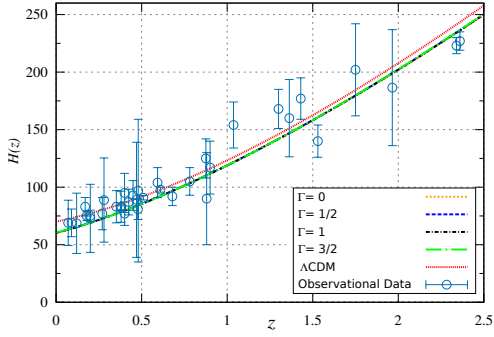

FIG. 3: Evolution of $H(z)$ as the function of the cosmic redshift for the thawing quintessence model of dark energy with the set of the different potentials.

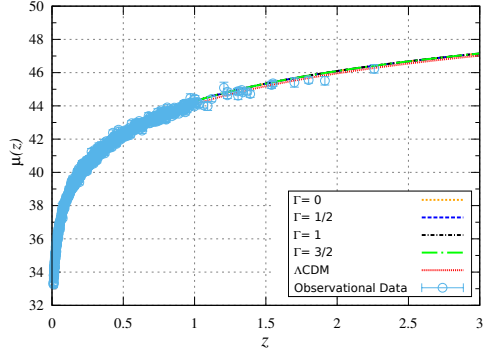

FIG. 4: Evolution of $\mu(z)$ as the function of the cosmic redshift for the thawing quintessence model of dark energy with the set of the different potentials.

Also, using a set of the observational data consisting of the SuperNovea type Ia (SNIa) [79], we can show the evolution of the distance modulus of SNIa ($\mu(z)$), as a function of cosmic redshift for the thawing quintessence model with the set of the different potentials. The distance modulus is defined as [80]

$$\mu(z) = 5\log_{10}[(1+z)\int_0^z \frac{dz}{E(z)}] + \mu_0, \quad (27)$$

where $\mu_0 = 42.384 - 5\log_{10}[h]$ is the current value of the distance modulus and $h = H_0/100$. In Figure (4) is shown that the evolution of $\mu(z)$ of the thawing quintessence model and the standard $\Lambda$CDM model are in the same behavior at low redshifts.

## III. GROWTH OF DENSITY PERTURBATIONS

In this sector, first, we study the linear growth of matter perturbations in the smooth thawing quintessence model with the set of the different potentials. Then, we compare the results of the smooth thawing quintessence model with the results of the standard $\Lambda$CDM model. For this purpose, we investigate the effects of dark energy on the linear growth of matter perturbations for clustered and smooth scenarios [81–101]. In the clustered scenario, the effective dark energy sound speed is approximately zero, $c_{eff}^2 \approx 0$, and dark energy perturbations will be able to grow in a identical process to matter perturbations and it occurs in sub-Hubble scales ($\delta_m \neq 0$, $\delta_\phi \neq 0$ and $c_{eff}^2 = 0$). In the smooth scenario, the effective dark energy sound speed is approximately one, $c_{eff}^2 \approx 1$, and dark energy perturbations can not grow on sub-Hubble scales and remains smooth ($\delta_m \neq 0$, $\delta_\phi = 0$ and $c_{eff}^2 = 1$).

Generally, it is considered that there are no dark energy perturbations in the thawing quintessence model on the cluster scales because in this model the mass of the field is tiny [61]. Hence, in this paper, we will investigate the thawing quintessence model with the set of the different potentials on the smooth scales and we just have the matter perturbations i.e. $\delta_m \neq 0$, $\delta_\phi = \mathbf{0}$ and $c_{eff}^2 = 1$. The basic and main equations that can be employed to describe the evolution of matter perturbations are as [48]

$$\dot{\theta}_m + H\theta_m - \frac{k^2\phi}{a} = 0, \quad (28)$$

$$\dot{\delta}_m + \frac{\theta_m}{a} = 0. \quad (29)$$

$k$ is the wave number of perturbations and $c_{eff}^2$ is the effective sound speed. Also, $\theta = \vec{\nabla} \cdot \vec{v}$ and $\delta_m$ are the velocity divergence and the dark matter pertuebation, respectively. One can also write the Poisson equation for the thawing quintessence model as:

$$-\frac{k^2}{a^2}\phi = \frac{3}{2}H^2\Big[\Omega_m\delta_m + (1+3c_{eff}^2)\Omega_\phi\delta_\phi\Big], \quad (30)$$

here $\delta_\phi$ is the dark energy pertuebations. Now, combining Eqs. (28), (30) and using Eq. (29) and assuming $c_{eff}^2 = 1$, for the smooth case, the differential equations of the evolution of dark matter perturbations can be written as:

$$\delta_m'' + A_m\delta_m' + B_m\delta_m = \frac{3}{2a^2}\Big(\Omega_m\delta_m + \Omega_\phi\delta_\phi\Big), \quad (31)$$

where the symbol $'$ denotes the derivative with respect to $a$. The coefficients of the equations are defined as follows:

$$A_m = \frac{3}{2a}\Big(1 - \omega_\phi \Omega_\phi\Big), \qquad (32)$$

$$B_m = 0. \qquad (33)$$

For the smooth case, we have $c_{eff}^2 = 1$, $\delta_\phi = 0$. Therefore, the differential equation of the evolution of dark matter perturbations can be obtained as: [45, 48, 102]

$$\delta_m'' + \frac{3}{2a}\Big(1 - \omega_\phi \Omega_\phi\Big)\delta_m' - \frac{3\Omega_{m_0}}{2a^5 E^2}\delta_m = 0. \qquad (34)$$

In order to numerically solve the Eq. (34), we are required appropriate initial condition [49, 103]

$$\delta_{mi}' = \frac{\delta_{mi}}{a_i}, \qquad (35)$$

and we apply the constraint as follows

$$a_i = 10^{-4} \qquad , \qquad \delta_{mi} = 8 \times 10^{-5}, \qquad (36)$$

Using the above conditions, we intend that the matter perturbations remain in the linear range region. Indeed, we set these values to guarantee that matter perturbations remain in the linear regime [45].
The linear growth factor to unity at the present time is defined [104]

$$D(a) = \frac{\delta_m(a)}{\delta_m(a=1)}. \qquad (37)$$

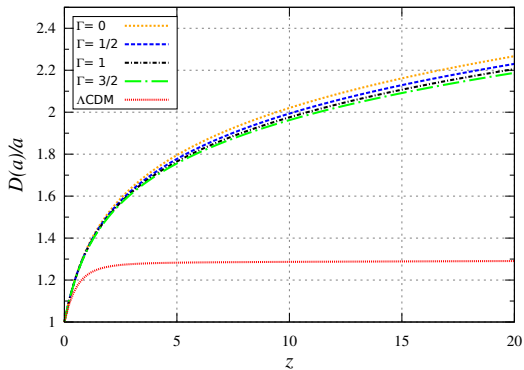

FIG. 5: Evolution of $D(a)/a$ as the function of the cosmic redshift for the smooth thawing quintessence model of dark energy.

Now, using the Eqs. (34), (35), (36) and (37), we have plotted $D(a)/a$ as the function of the cosmic redshift for the smooth thawing quintessence model with the set of the different potentials and the standard $\Lambda$CDM model in Figure (5).
In this Figure, we can observe the evolution of the linear growth factor to unity at the present time as a function of cosmic redshift, $z$, for the standard $\Lambda$CDM model and the smooth thawing quintessence model of dark energy with the set of the different potentials. In general, we can find out that the amplitude of the linear growth factor of the matter perturbations for the smooth thawing quintessence model of dark energy with the set of the different potentials reduces as $\Gamma$. We can see that the amplitude of the linear growth factor of the matter perturbations is larger than the standard $\Lambda$CDM model at high redshifts. In the standard $\Lambda$CDM model, for $z > 2$, we can see the amplitude of the linear growth factor of the matter perturbations reaches a plateau which this constant value implies that the impact of the cosmological constant on the growth of cosmic structures is insignificant. However, it is not the case for the smooth thawing quintessence model of dark energy and in this case, it seems to evolve even at $z > 20$. This behavior of the linear growth factor can be interpreted as a small effect but non-negligible effect of the dark energy component on the growth of perturbations. Also, at low redshifts, the effect of the dark energy component on the growth of the perturbations is significant. Hence we can conclude that at high redshifts the dark energy component reduces the growth of the cosmic structures.

Now, we can focus on the analysis of the growth index of matter perturbations, $\gamma$, as follows [105]

$$F(a) = \frac{d\ln\delta_m}{d\ln a} \simeq \Omega_m^\gamma. \qquad (38)$$

If we combine Eqs. (28), (29) and (30), then we can obtain [94]

$$a^2 \delta_m'' + a\Big(3 + \frac{\dot{H}}{H^2}\Big)\delta_m' = \frac{3}{2}\Omega_m \mu(a), \qquad (39)$$

where

$$\frac{\dot{H}}{H^2} = \frac{d\ln H}{d\ln a} = -\frac{3}{2} - \frac{3}{2}\omega_\phi(a)\Omega_\phi(a), \qquad (40)$$

where for the smooth dark energy model, the quantity $\mu(a)$ is 1 [106]. Furthermore, substituting Eqs. (38) and (40) in Eq. (39), we can obtain [106]

$$-(1+z)\frac{d\gamma}{dz}\ln(\Omega_m) + \Omega_m^\gamma + 3\omega_\phi\Omega_\phi\Big(\gamma - \frac{1}{2}\Big)$$
$$+ \frac{1}{2} = \frac{3}{2}\Omega_m^{1-\gamma}\mu. \qquad (41)$$

Now, for the growth index, we can use the following phenomenological parameterisation [106, 107]

$$\gamma(a) = \gamma_0 + \gamma_1[1 - a(z)]. \qquad (42)$$

Using Eqs. (41) and (42) at the present time $z = 0$, we arrive at [107]

$$\gamma_1 = \frac{\Omega_{m0}^{\gamma_0} + 3\omega_{\phi 0}(\gamma_0 - \frac{1}{2})\Omega_{\phi 0} + \frac{1}{2} - \frac{3}{2}\Omega_{m0}^{1-\gamma_0}\mu_0}{\ln \Omega_{m0}}, \qquad (43)$$

where $\mu_0 = \mu(z=0)$ and $\omega_{\phi 0} = \omega_\phi(z=0)$. In order to predict the evolution of the growth index in the smooth thawing quintessence model of dark energy, it needs to estimate the value of $\gamma_0$. For this purpose, we can utilize $\gamma_\infty \simeq \gamma_0 + \gamma_1$ which $\gamma_\infty$ is given by [105]

$$\gamma_\infty = \frac{3(M_0 + M_1) - 2(H_1 + N_1)}{2 + 2X_1 + 3M_0}, \qquad (44)$$

where

$$M_0 = \mu|_{\omega=0}, \quad M_1 = \frac{d\mu}{d\omega}|_{\omega=0},$$
$$N_1 = 0, \quad H_1 = -\frac{X_1}{2} = \frac{3}{2}\omega_\phi(a)|_{\omega=0}, \qquad (45)$$

here $\omega = \ln \Omega_m(a)$. Using Eq. (45), for the smooth thawing quintessence model of dark energy, we can obtain $\{M_0, M_1, H_1, X_1\} = \{1, 0, \frac{3}{2}\omega_\phi, -3\omega_\phi\}$. Substituting the values of $\{M_0, M_1, H_1, X_1\}$ in Eq. (44), we obtain

$$\gamma_\infty = \frac{3(\omega_\phi - 1)}{6\omega_\phi - 5}. \qquad (46)$$

Using Eqs. (43), (46) and $\gamma_0 \simeq \gamma_\infty - \gamma_1$ for the smooth thawing quintessence model of dark energy with the set of the different potentials, we obtain

$$\Gamma = 0 \ \& \ (\gamma_0, \gamma_1, \gamma_\infty) = (1.131, -0.579, 0.5523),$$
$$\Gamma = 1/2 \ \& \ (\gamma_0, \gamma_1, \gamma_\infty) = (1.130, -0.580, 0.550),$$
$$\Gamma = 1 \ \& \ (\gamma_0, \gamma_1, \gamma_\infty) = (1.1294, -0.5794, 0.5496),$$
$$\Gamma = 3/2 \ \& \ (\gamma_0, \gamma_1, \gamma_\infty) = (1.129, -0.580, 0.549).$$
$$(47)$$

Now, for the smooth thawing quintessence model with the set of the different potentials and the standard $\Lambda$CDM model we investigate the growth rate, $f(a)$, as function of redshift. The growth rate is defined [104]

$$f(a) = \frac{d \ln D(a)}{d \ln a}, \qquad (48)$$

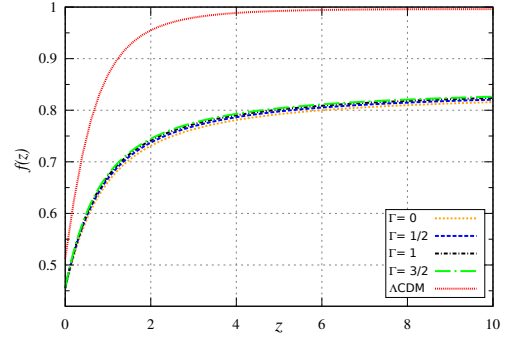

(a)

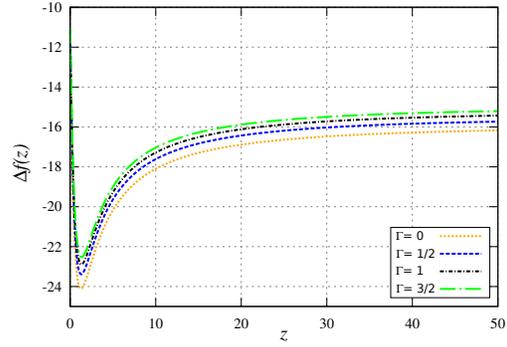

(b)

FIG. 6: (a) Evolution of $f(z) - z$ for the smooth thawing quintessence model of dark energy.
(b) Evolution of $\Delta f(z) - z$ for the smooth thawing quintessence model of dark energy.

Here, $D(a)$ and $a$ are the linear growth factor and scale factor, respectively. Moreover, we compute the fractional difference of the growth rate with respect to the standard $\Lambda$CDM model [104]

$$\Delta f(\%) = \left(\frac{f(a)_{model} - f(a)_{\Lambda CDM}}{f(a)_{\Lambda CDM}}\right) \times 100, \qquad (49)$$

where $f(a)_{model}$ is the growth rate for the smooth thawing quintessence model with the set of the different potentials and $f(a)_{\Lambda CDM}$ is the growth rate for the standard $\Lambda$CDM model. We have plotted $f(a)$ and $\Delta f(\%)$ as function of redshift $z$ for the smooth thawing quintessence model of dark energy with the set of the different potentials in Figures (6-a) and (6-b).

In Figure (6-a), we can figure out that the amplitude of the growth rate increases as $\Gamma$. In this case, we can see that the amplitude of the growth rate of the matter perturbations is smaller than the standard

ΛCDM model. Also, at high $z$, we can see in Figure (6-a) $f(z)_{\Lambda CDM} \sim 1$ and at $z > 6$ the amplitude of the growth rate of the matter perturbations reaches a constant value, $\sim 1$. But for the smooth thawing quintessence model of dark energy with the set of the different potentials $f(z) \sim 0.8$ at high $z$ and the amplitude of the growth rate of the matter perturbations reaches a constant value, $\sim 0.8$. However, at high redshifts, the evolution of the growth rate implies the effect of all models on the growth rate of the cosmic structures is not important. In other words, at low redshifts, the effect of the dark energy component on the growth rate of the perturbations is significant and non-negligible. Hence, the dark energy component stops the growth of the cosmic structures.

In Figure (6-b), we have plotted $\Delta f(\%)$ as function of redshift $z$ for the smooth thawing quintessence model of dark energy with the set of the different potentials. For $0 \leqslant z \leqslant 50$, we can obtain:

$$\Delta f \sim [-16.1\%, -11.4\%] \quad \text{for} \quad \Gamma = 0,$$
$$\Delta f \sim [-15.7\%, -11.2\%] \quad \text{for} \quad \Gamma = 1/2,$$
$$\Delta f \sim [-15.4\%, -11.1\%] \quad \text{for} \quad \Gamma = 1,$$
$$\Delta f \sim [-15.2\%, -11.02\%] \quad \text{for} \quad \Gamma = 3/2.$$

Finally, in this section, the evolution of $f(z)\sigma_8(z)$ with respect to the redshifts for the standard ΛCDM model, the observational data [70, 108–121], the smooth thawing quintessence model of dark energy with the set of the different potentials are shown in Figure (7) for the smooth perturbations. In this case, we can see that all of the models have a behavior relatively in agreement with observational data at low redshifts.

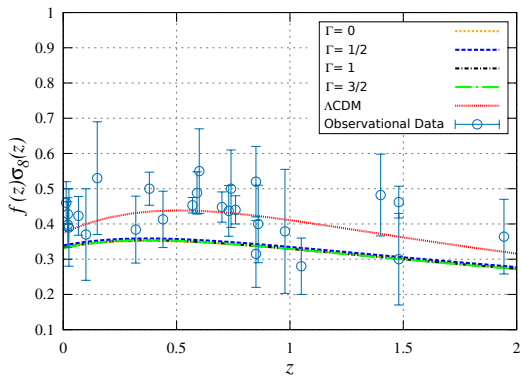

FIG. 7: Evolution of $f(z)\sigma_8(z)$ as a function of the cosmic redshift for the standard ΛCDM model, the observational data and the smooth thawing quintessence model of dark energy with the set of the different potentials.

## IV. OBSERVATIONAL CONSTRAINTS ON THE THAWING DARK ENERGY MODEL

In this section, first, we carry out an overall likelihood analysis utilizing the latest available observational data on the thawing quintessence model with two different cases: the free parameter $\Gamma$ and the constant parameter $\Gamma = 1$. Next, using the growth rate data, we investigate these models at the smooth perturbation level. Then, using the well-known information criteria, namely Akaike and Bayesian information, we study the agreement of these models with the latest observational data sets at the background and the smooth perturbation levels. Finally, the acquired results with the results of the standard ΛCDM model are perused.

The overall likelihood function is given by the product of the individual likelihoods [104]:

$$\mathcal{L}_{tot}(\mathbf{p}) = \mathcal{L}_{SN} \times \mathcal{L}_{BAO} \times \mathcal{L}_{CMB} \times \mathcal{L}_H \times \mathcal{L}_{BBN} \times \mathcal{L}_{GR}, \quad (50)$$

and the total chi-square function, $\chi^2_{tot}$, is described as [104]

$$\chi_{tot}(\mathbf{p}) = \chi_{SN} + \chi_{BAO} + \chi_{CMB} + \chi_H + \chi_{BBN} + \chi_{GR}. \quad (51)$$

We define $\mathbf{p}$ as the statistical vector of free parameters and would like to put constraints on them. As we mentioned before, we are going to study the thawing quintessence model with two different $\Gamma$ cases. For the free parameter $\Gamma$ case, $\{\Gamma, \Omega_{DM}, \Omega_B, h, \sigma_8\}$ are assumed as the free parameters of the thawing quintessence model, in which $\sigma_8(z)$ is the mass variance of the overdensity on the scale of $R_8 = 8h^{-1}Mpc$ and $h$ is determined as $h = H_0/100$. In order to investigate this model at the background level, we consider $\{\Gamma, \Omega_{DM}, \Omega_B, h\}$ as free parameters. In this analysis, we fix the radiation density (photons and relativistic neutrinos) as $\Omega_r = 2.469 \times 10^{-5}h^{-2}(1.6903)$, therefore, it is not a free parameter [122]. For the constant parameter $\Gamma = 1$ case, $\{\Omega_{DM}, \Omega_B, h, \sigma_8\}$ are assumed as the free parameters of the thawing quintessence model. In order to investigate this model at the background level, we consider $\{\Omega_{DM}, \Omega_B, h\}$ as free parameters. In Eqs. (50) and (51), we apply the various observational data consisting of the SuperNovae type Ia (SN) [79], the Baryonic Acoustic Oscillation (BAO) [4, 123–125], the Cosmic Microwave Background (CMB) [103, 126, 127], the Hubble parameter (H) [68–78], the Big Bang Nucleosynthesis (BBN) [128, 129] and the Growth Rate data (GR) [70, 108–121]. In this work, we use 1048 the SnIa data [79], 36 the Hubble data [68–77], 26 the growth rate data [70, 108–121], and the BAO data based on 6 distinct measurements of the baryon acoustic scale (see Table 1 of [103]).

In our study, to acquire the best fit of the free parameters and their confidence regions with data for the thawing quintessence and the standard $\Lambda$CDM models, we utilize the Markov Chain Monte Carlo (MCMC) procedure at the background and the smooth perturbation levels. The MCMC procedure has been comprehensively investigated in the literature that we can see them in [103, 104, 126, 130, 131]. In Tables (I), (II), (III) and (IV), we have obtained the best-fit values for the thawing quintessence scalar field and the standard $\Lambda$CDM models at the background and smooth perturbation levels. These tables show that in the thawing quintessence scalar field model with the free parameter $\Gamma$ and the constant parameter $\Gamma = 1$ cases, the contributions of dark matter and baryon have substantial differences with the standard $\Lambda$CDM model at the background and smooth perturbation levels. At both levels, the contribution of $H_0$ parameter has a relatively difference with the standard $\Lambda$CDM model. It is needed to mention that the value of $\sigma_8$ in Table (II) is very low for the thawing quintessence scalar field model with the constant parameter $\Gamma = 1$ case. As it is mentioned before, the thawing quintessence scalar field model depends on its potential. Therefore, according to Eq. (25), it may be because of that kind of selection potential i.e. $V = e^{\alpha \phi}$. Therefore, the choice of exponential potential reduces the speed of the linear growth factor and $\sigma_8$.

In order to study the statistics of these models, we utilize two famous information criteria: one is the Akaike Information Criteria (AIC) and another is the Bayesian Information Criteria (BIC). The AIC is determined as[132]

$$AIC = -2\ln \mathcal{L}_{max} + 2k + \frac{2k(k+1)}{N-k-1}. \quad (52)$$

and The BIC is given by [133]

$$BIC = -2\ln \mathcal{L}_{max} + k \ln N. \quad (53)$$

If we have $N/k \gg 1$, the AIC, Eq. (52), is could be reduced [132]

$$AIC = -2\ln \mathcal{L}_{max} + 2k, \quad (54)$$

where $k$ is the number of the free parameters, $N$ is the number of data and $\mathcal{L}_{max}$ is the maximum likelihood. In both criteria, the best model will be the one with the lowest value. The important statistical results, $AIC$, $BIC$ and $\chi_i^2$, are demonstrated in Tables (V) and (VI) for the thawing quintessence model scalar field model with two different $\Gamma$ cases and the standard $\Lambda$CDM model at the background and smooth perturbation levels. These Tables show that the thawing quintessence scalar field model with two different $\Gamma$ cases at both levels has significant differences with the standard $\Lambda$CDM model. We guess that the major difference in the amounts of the dark matter, the baryon, and the Hubble parameters in the thawing quintessence model with the standard $\Lambda$CDM model, be the reason for those.

Now, for studying the statistical representation of the models, we can use $\Delta AIC = AIC_{model} - AIC_{min}$ and $\Delta BIC = BIC_{model} - BIC_{min}$ which the subscript $min$ is considered as the standard $\Lambda$CDM model. For the background and smooth levels, we can obtain the following results:

• For the thawing quintessence model with the free parameter $\Gamma$ case at the background level: $\Delta AIC = 243$ and $\Delta BIC = 248.042$. According to these obtained values, we have $\Delta AIC > 10$ and $\Delta BIC > 10$. These results are very strong evidence for the incompatibility of this model.

• For the thawing quintessence model with the constant parameter $\Gamma = 1$ case at the background level: $\Delta AIC = 204.5$ and $\Delta BIC = 204.532$. According to these obtained values, we have $\Delta AIC > 10$ and $\Delta BIC > 10$. These results are very strong evidence for the incompatibility of this model.

• For the thawing quintessence model with the free parameter $\Gamma$ case at the smooth level: $\Delta AIC = 247.3$ and $\Delta BIC = 252.353$. According to these obtained values, we have $\Delta AIC > 10$ and $\Delta BIC > 10$. These results are very strong evidence for the incompatibility of the two models.

• For the thawing quintessence model with the constant parameter $\Gamma = 1$ case at the smooth level: $\Delta AIC = 921.52$ and $\Delta BIC = 921.563$. According to these obtained values, we have $\Delta AIC > 10$ and $\Delta BIC > 10$. These results are very strong evidence for the incompatibility of the two models.

Eventually, at the background and smooth perturbation levels, the contours of the $1\sigma$, $2\sigma$ and $3\sigma$ confidence levels are shown for the thawing quintessence scalar field model with two different $\Gamma$ cases and the standard $\Lambda$CDM model in Figures (8) and (9). As you can see in these Figures, we plot the contours of the confidence levels based on a set of observational data consisting of SNIa, BAO, CMB, H(z), and BBN. We also plot the contours of the confidence levels based on a set of observational data including SNIa, BAO, CMB, H(z), BBN, and $f(z)\sigma_8(z)$ [3, 49, 50, 52, 53, 69, 70, 72–77, 79, 102–104, 108–120, 131, 134–150].

TABLE I: Best-fit values for the standard $\Lambda$CDM model and the thawing quintessence scalar field model with the free parameter $\Gamma$ at the smooth perturbation level.

| Model | Quintessence | $\Lambda$CDM |
|---|---|---|
| $\Omega_{DM}$ | $0.1892^{+0.0047,+0.0079,+0.010}_{-0.0039,-0.0086,-0.011}$ | $0.2411^{+0.0036,+0.0069,+0.0083}_{-0.0030,-0.0073,-0.0085}$ |
| $\Omega_B$ | $0.0621^{+0.0018,+0.0035,+0.0041}_{-0.0018,-0.0034,-0.0043}$ | $0.0462^{+0.00047,+0.00094,+0.0016}_{-0.00043,-0.00095,-0.0011}$ |
| $H_0$ | $65.5^{+0.0052,+0.013,+0.017}_{-0.0075,-0.011,-0.013}$ | $69.61^{+0.0024,+0.0058,+0.0069}_{-0.0029,-0.0055,-0.0067}$ |
| $\sigma_8$ | $0.653^{+0.053,+0.10,+0.14}_{-0.053,-0.098,-0.13}$ | $0.8055^{+0.0053,+0.011,+0.013}_{-0.0053,-0.011,-0.014}$ |
| $\Gamma$ | $0.963^{+0.020,+0.037,+0.054}_{-0.020,-0.036,-0.051}$ | $------$ |

TABLE II: Best-fit values for the standard $\Lambda$CDM model and the thawing quintessence scalar field model with the constant parameter $\Gamma = 1$ at the smooth perturbation level.

| Model | Quintessence | $\Lambda$CDM |
|---|---|---|
| $\Omega_{DM}$ | $0.1820^{+0.0021,+0.0038,+0.0059}_{-0.0021,-0.0039,-0.0055}$ | $0.2411^{+0.0036,+0.0069,+0.0083}_{-0.0030,-0.0073,-0.0085}$ |
| $\Omega_B$ | $0.0567^{+0.0014,+0.0023,+0.0029}_{-0.0011,-0.0025,-0.0033}$ | $0.0462^{+0.00047,+0.00094,+0.0016}_{-0.00043,-0.00095,-0.0011}$ |
| $H_0$ | $67.26^{+0.0022,+0.0066,+0.0087}_{-0.0034,-0.0053,-0.0064}$ | $69.61^{+0.0024,+0.0058,+0.0069}_{-0.0029,-0.0055,-0.0067}$ |
| $\sigma_8$ | $0.574^{+0.036,+0.082,+0.081}_{-0.048,-0.072,-0.073}$ | $0.8055^{+0.0053,+0.011,+0.013}_{-0.0053,-0.011,-0.014}$ |

TABLE III: Best-fit values for the standard $\Lambda$CDM model and the thawing quintessence scalar field model with the free parameter $\Gamma$ at the background level.

| Model | Quintessence | $\Lambda$CDM |
|---|---|---|
| $\Omega_{DM}$ | $0.1745^{+0.0032,+0.0058,+0.0068}_{-0.0032,-0.0063,-0.0073}$ | $0.2386^{+0.0048,+0.0077,+0.0096}_{-0.0072,-0.013,-0.018}$ |
| $\Omega_B$ | $0.0574^{+0.0016,+0.0029,+0.0035}_{-0.0015,-0.0030,-0.0038}$ | $0.0460^{+0.00051,+0.00094,+0.0013}_{-0.0039,-0.0064,-0.0079}$ |
| $H_0$ | $68.2^{+0.0049,+0.013,+0.015}_{-0.0060,-0.010,-0.012}$ | $69.6^{+0.0071,+0.014,+0.017}_{-0.0071,-0.015,-0.018}$ |
| $\Gamma$ | $98.61^{+0.080,+0.17,+0.18}_{-0.10,-0.16,-0.16}$ | $------$ |
| $\omega_\phi(z=0)$ | $-0.99606$ | $-1.00$ |
| $\Omega_\phi(z=0)$ | $0.70$ | $0.70$ |

TABLE IV: Best-fit values for the standard $\Lambda$CDM model and the thawing quintessence scalar field model with the constant parameter $\Gamma = 1$ at the background level.

| Model | Quintessence | $\Lambda$CDM |
|---|---|---|
| $\Omega_{DM}$ | $0.191^{+0.0049,+0.0099,+0.011}_{-0.0049,-0.10,-0.014}$ | $0.2386^{+0.0048,+0.0077,+0.0096}_{-0.0072,-0.013,-0.018}$ |
| $\Omega_B$ | $0.0631^{+0.0019,+0.0038,+0.0048}_{-0.0019,-0.0038,-0.0046}$ | $0.0460^{+0.00051,+0.00094,+0.0013}_{-0.0039,-0.0064,-0.0079}$ |
| $H_0$ | $65.11^{+0.0076,+0.017,+0.023}_{-0.0085,-0.016,-0.017}$ | $69.6^{+0.0071,+0.014,+0.017}_{-0.0071,-0.015,-0.018}$ |
| $\omega_\phi(z=0)$ | $-0.84503$ | $-1.00$ |
| $\Omega_\phi(z=0)$ | $0.70$ | $0.70$ |

TABLE V: Best-fit values for the standard $\Lambda$CDM model and the thawing quintessence scalar field model with the free parameter $\Gamma$ at the background and the smooth perturbation levels.

| Model | Quintessence | $\Lambda$CDM | level |
|---|---|---|---|
| $\chi^2$ | $1299.3^{+0.77,+3.7,+6.1}_{-2.0,-2.8,-3.2}$ | $1058.3^{+0.60,+6.5,+17.0}_{-3.1,-3.9,-4.3}$ | Background |
| $AIC$ | $1307.3$ | $1064.3$ | Background |
| $BIC$ | $1327.34$ | $1079.298$ | Background |
| $\chi^2$ | $1316.8^{+1.4,+5.6,+8.5}_{-2.8,-4.2,-4.9}$ | $1071.5^{+0.76,+5.1,+8.4}_{-3.0,-3.8,-4.3}$ | smooth perturbation |
| $AIC$ | $1326.8$ | $1079.5$ | smooth perturbation |
| $BIC$ | $1351.85$ | $1099.497$ | smooth perturbation |

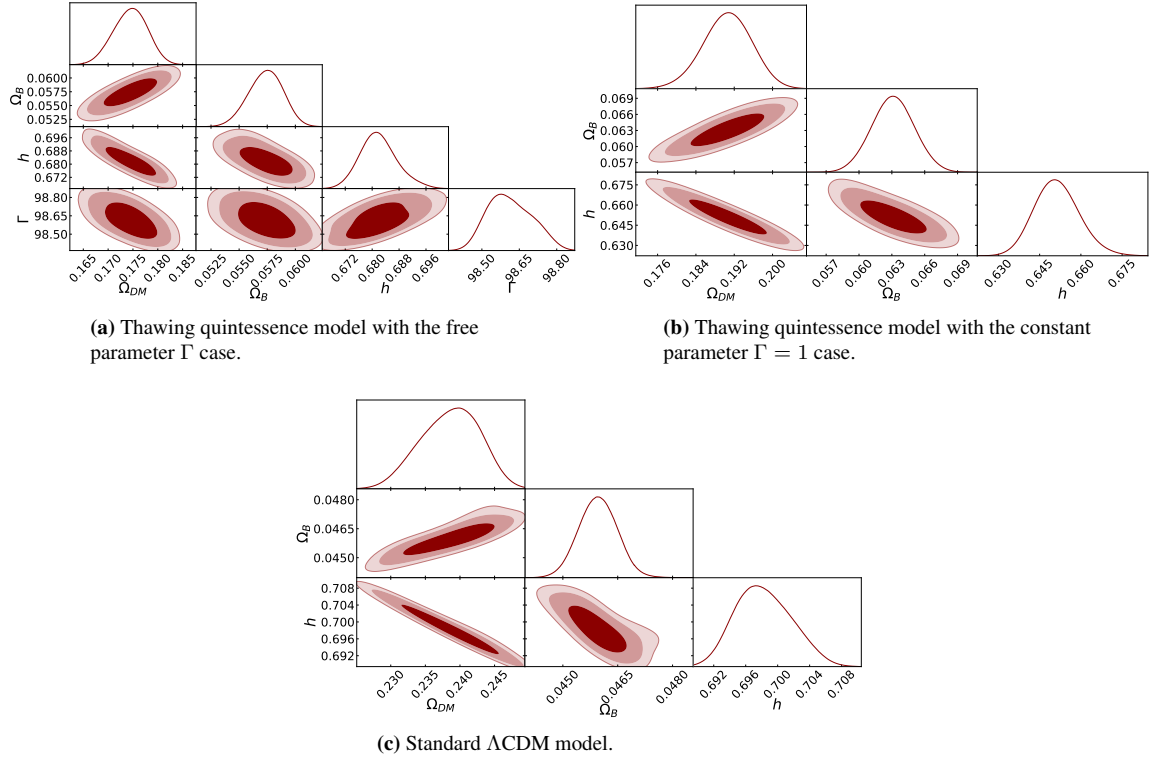

(a) Thawing quintessence model with the free parameter $\Gamma$ case.

(b) Thawing quintessence model with the constant parameter $\Gamma = 1$ case.

(c) Standard $\Lambda$CDM model.

FIG. 8: $1\sigma$, $2\sigma$ and $3\sigma$ confidence contours for the thawing quintessence scalar field and the standard $\Lambda$CDM models at the background level.

TABLE VI: Best-fit values for the standard $\Lambda$CDM model and the thawing quintessence scalar field model with the constant parameter $\Gamma = 1$ at the background and the smooth perturbation levels.

| Model | Quintessence | $\Lambda$CDM | level |
|---|---|---|---|
| $\chi^2$ | $1262.8^{+0.99,+4.6,+9.1}_{-2.4,-3.4,-3.7}$ | $1058.3^{+0.60,+6.5,+17.0}_{-3.1,-3.9,-4.3}$ | Background |
| $AIC$ | 1268.8 | 1064.3 | Background |
| $BIC$ | 1283.83 | 1079.298 | Background |
| $\chi^2$ | $1993.02^{+0.46,+6.0,+14.0}_{-2.5,-3.1,-3.4}$ | $1071.5^{+0.76,+5.1,+8.4}_{-3.0,-3.8,-4.3}$ | smooth perturbation |
| $AIC$ | 2001.02 | 1079.5 | smooth perturbation |
| $BIC$ | 2021.06 | 1099.497 | smooth perturbation |

## V. CONCLUSION

In this work, using the various observational data consisting of the SuperNovae type Ia (SNIa), the Baryonic Acoustic Oscillation (BAO), the Cosmic Microwave Background (CMB), the Hubble parameter, the Big Bang Nucleosynthesis (BBN) and the Growth Rate data, the thawing quintessence scalar field model is constrained. The evolution of equation of state parameter, $\omega_\phi(z)$, the density parameter, $\Omega_\phi(z)$, the dimensionless Hubble parameter, $E(z)$, the deceleration parameter, $q(z)$ and the distance modulus of SNIa, $\mu(z)$ as a function of redshift are investigated for the thawing quintessence scalar field and the standard $\Lambda$CDM models.

As the thawing quintessence scalar field model depends on its potential, so it is chosen the set of the different potentials as $V = \phi^\alpha$ and $e^{\alpha\phi}$ which $V = \phi^\alpha$ is included $V_1 = \phi, V_2 = \phi^2, V_3 = \phi^{-2}$ and $e^{\alpha\phi}$ which is included $V_4 = e^\phi, V_5 = e^{-\phi}$. It is demonstrated that the evolution of the EoS parameter of the thawing quintessence scalar field model with the set of the different potentials has different behavior when their evolution approaches the present time and all models approach $-1$ at high redshift whereas the evolution of the density parameter and the dimensionless Hubble parameter of this model with the set of the different potentials have the almost same behavior in all times but their evolution behaves differently from the standard $\Lambda$CDM model.

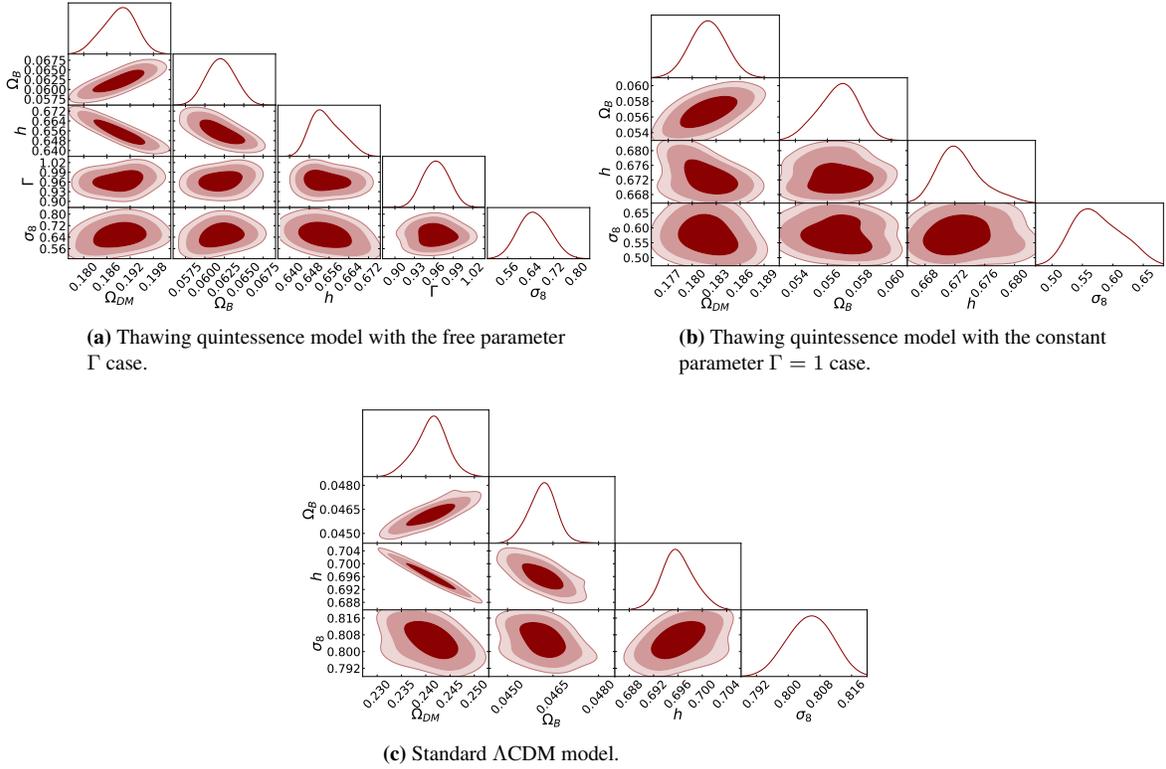

FIG. 9: $1\sigma$, $2\sigma$ and $3\sigma$ confidence contours for the thawing quintessence scalar field and the standard $\Lambda$CDM models at the smooth level.

Comparing the evolution of the Hubble parameter and the distance modulus of SNIa have a same behavior with the standard $\Lambda$CDM model at low redshift for the thawing quintessence scalar field model with the set of the different potentials. We have also discovered that the thawing quintessence scalar field model with the set of the different potentials warrants the acceleration of the universe's expansion and the evolution of the deceleration parameter of them enter to accelerating phase earlier than the standard $\Lambda$CDM model. This is the reason that in these models, the evolution of the Hubble parameter is faster than the evolution of the Hubble parameter in the standard $\Lambda$CDM model.

The amplitude of the linear growth factor of the matter perturbations for the thawing quintessence scalar field models is larger than the amplitude of the standard $\Lambda$CDM model at high redshift and the effect of the dark energy component on linear growth of perturbations in all models at low redshifts is non-negligible. Therefore, the presence of the dark energy component hinders the growth of cosmic structures as we expect. This behavior of the linear growth factor can be interpreted as a small effect but non-negligible effect of the dark energy component on the growth of perturbations. Also, at low redshifts, the effect of the dark energy component on the growth of the perturbations is significant. Hence, we can conclude that at high redshifts the dark energy component reduces the growth of the cosmic structures. Moreover, we have compared the observational data of $f(z)\sigma_8(z)$ with the standard $\Lambda$CDM model and the smooth thawing quintessence scalar field model of dark energy with the set of the different potentials and we have obtained that all of the models have a behavior relatively in agreement with the observational data at low redshifts.

In our study, to acquire the best fit of the free parameters and their confidence regions with data for the thawing quintessence scalar field and the standard ΛCDM models, we have utilized the Markov Chain Monte Carlo (MCMC) procedure at the background and the smooth perturbation levels. The important results were represented in Tables (I), (II), (III), (IV), (V) and (VI). These tables show that in the thawing quintessence scalar field model with the free parameter $\Gamma$ and the constant parameter $\Gamma = 1$, the contributions of dark matter and baryon have substantial differences with the standard ΛCDM model at the background and smooth perturbation levels. At both levels, the contribution of $H_0$ parameter has a relatively large difference with the standard ΛCDM model. These values predict that the thawing quintessence scalar field model does not have a good agreement with the standard ΛCDM model at the background and smooth perturbation levels. In order to study the statistics of these models, we have utilized the Akaike Information Criteria and the Bayesian Information Criteria. It has displayed that these models with the free parameter $\Gamma$ and the constant parameter $\Gamma = 1$ are inconsistent with the observational data at both levels. Hence, we should be looking for consistent potentials with observational data.

### ACKNOWLEDGMENTS

The work of KB was supported in part by the JSPS KAKENHI Grant Number JP21K03547 and 23KF0008.

[1] D. J. Eisenstein *et al.* (SDSS), Detection of the Baryon Acoustic Peak in the Large-Scale Correlation Function of SDSS Luminous Red Galaxies, Astrophys. J. **633**, 560 (2005), arXiv:astro-ph/0501171 [astro-ph].

[2] H.-J. Seo and D. J. Eisenstein, Baryonic acoustic oscillations in simulated galaxy redshift surveys, Astrophys. J. **633**, 575 (2005), arXiv:astro-ph/0507338 [astro-ph].

[3] C. Blake *et al.*, The WiggleZ Dark Energy Survey: the growth rate of cosmic structure since redshift z=0.9, Mon. Not. Roy. Astron. Soc. **415**, 2876 (2011), arXiv:1104.2948 [astro-ph.CO].

[4] C. Blake *et al.*, The WiggleZ Dark Energy Survey: mapping the distance-redshift relation with baryon acoustic oscillations, Mon. Not. Roy. Astron. Soc. **418**, 1707 (2011), arXiv:1108.2635 [astro-ph.CO].

[5] E. Komatsu *et al.* (WMAP), Five-Year Wilkinson Microwave Anisotropy Probe (WMAP) Observations: Cosmological Interpretation, Astrophys. J. Suppl. **180**, 330 (2009), arXiv:0803.0547 [astro-ph].

[6] N. Jarosik *et al.*, Seven-Year Wilkinson Microwave Anisotropy Probe (WMAP) Observations: Sky Maps, Systematic Errors, and Basic Results, Astrophys. J. Suppl. **192**, 14 (2011), arXiv:1001.4744 [astro-ph.CO].

[7] E. Hawkins *et al.*, The 2dF Galaxy Redshift Survey: Correlation functions, peculiar velocities and the matter density of the universe, Mon. Not. Roy. Astron. Soc. **346**, 78 (2003), arXiv:astro-ph/0212375 [astro-ph].

[8] M. Tegmark *et al.* (SDSS), Cosmological parameters from SDSS and WMAP, Phys. Rev. **D69**, 103501 (2004), arXiv:astro-ph/0310723 [astro-ph].

[9] S. Cole *et al.* (2dFGRS), The 2dF Galaxy Redshift Survey: Power-spectrum analysis of the final dataset and cosmological implications, Mon. Not. Roy. Astron. Soc. **362**, 505 (2005), arXiv:astro-ph/0501174 [astro-ph].

[10] A. G. Riess *et al.* (Supernova Search Team), Observational evidence from supernovae for an accelerating universe and a cosmological constant, Astron. J. **116**, 1009 (1998), arXiv:astro-ph/9805201.

[11] S. Perlmutter *et al.* (Supernova Cosmology Project), Measurements of $\Omega$ and $\Lambda$ from 42 high redshift supernovae, Astrophys. J. **517**, 565 (1999), arXiv:astro-ph/9812133.

[12] M. Kowalski *et al.* (Supernova Cosmology Project), Improved Cosmological Constraints from New, Old and Combined Supernova Datasets, Astrophys. J. **686**, 749 (2008), arXiv:0804.4142 [astro-ph].

[13] S. Weinberg, The cosmological constant problem, Reviews of Modern Physics **61**, 1 (1989).

[14] S. M. Carroll, The Cosmological constant, Living Rev. Rel. **4**, 1 (2001), arXiv:astro-ph/0004075 [astro-ph].

[15] E. J. Copeland, M. Sami, and S. Tsujikawa, Dynamics of dark energy, Int. J. Mod. Phys. **D15**, 1753 (2006), arXiv:hep-th/0603057 [hep-th].

[16] T. Padmanabhan, Dark energy and gravity, Gen. Rel. Grav. **40**, 529 (2008), arXiv:0705.2533 [gr-qc].

[17] R. Durrer and R. Maartens, Dark Energy and Dark Gravity, Gen. Rel. Grav. **40**, 301 (2008), arXiv:0711.0077 [astro-ph].

[18] K. Bamba, S. Capozziello, S. Nojiri, and S. D. Odintsov, Dark energy cosmology: the equivalent description via different theoretical models and cosmography tests, Astrophys. Space Sci. **342**, 155 (2012), arXiv:1205.3421 [gr-qc].

[19] S. Capozziello and M. De Laurentis, Extended Theories of Gravity, Phys. Rept. **509**, 167 (2011), arXiv:1108.6266 [gr-qc].

[20] S. Nojiri and S. D. Odintsov, Unified cosmic history in modified gravity: from F(R) theory to Lorentz non-invariant models, Phys. Rept. **505**, 59 (2011), arXiv:1011.0544 [gr-qc].

[21] S. Nojiri, S. D. Odintsov, and V. K. Oikonomou, Modified Gravity Theories on a Nutshell: Inflation, Bounce and Late-time Evolution, Phys. Rept. **692**, 1 (2017), arXiv:1705.11098 [gr-qc].

[22] T. P. Sotiriou and V. Faraoni, f(R) Theories Of Gravity, Rev. Mod. Phys. **82**, 451 (2010), arXiv:0805.1726 [gr-qc].

[23] A. De Felice and S. Tsujikawa, f(R) theories, Living Rev. Rel. **13**, 3 (2010), arXiv:1002.4928 [gr-qc].

[24] K. Bamba and S. D. Odintsov, Inflationary cosmology in modified gravity theories, Symmetry **7**, 220 (2015), arXiv:1503.00442 [hep-th].

[25] Y.-F. Cai, S. Capozziello, M. De Laurentis, and E. N. Saridakis, f(T) teleparallel gravity and cosmology, Rept. Prog. Phys. **79**, 106901 (2016), arXiv:1511.07586 [gr-qc].

[26] A. Sen, Rolling tachyon, JHEP **04**, 048, arXiv:hep-th/0203211 [hep-th].

[27] T. Padmanabhan and T. R. Choudhury, Can the clustered dark matter and the smooth dark energy arise from the same scalar field?, Phys. Rev. **D66**, 081301 (2002), arXiv:hep-th/0205055 [hep-th].

[28] C. Wetterich, Cosmology and the Fate of Dilatation Symmetry, Nucl. Phys. **B302**, 668 (1988), arXiv:1711.03844 [hep-th].

[29] B. Ratra and P. J. E. Peebles, Cosmological consequences of a rolling homogeneous scalar field, Phys. Rev. D **37**, 3406 (1988).

[30] R. R. Caldwell, A Phantom menace?, Phys. Lett. **B545**, 23 (2002), arXiv:astro-ph/9908168 [astro-ph].

[31] S. Nojiri and S. D. Odintsov, Quantum de Sitter cosmology and phantom matter, Phys. Lett. **B562**, 147 (2003), arXiv:hep-th/0303117 [hep-th].

[32] S. Nojiri and S. D. Odintsov, DeSitter brane universe induced by phantom and quantum effects, Phys. Lett. **B565**, 1 (2003), arXiv:hep-th/0304131 [hep-th].

[33] E. Elizalde, S. Nojiri, and S. D. Odintsov, Late-time cosmology in (phantom) scalar-tensor theory: Dark energy and the cosmic speed-up, Phys. Rev. **D70**, 043539 (2004), arXiv:hep-th/0405034 [hep-th].

[34] S. Nojiri, S. D. Odintsov, and S. Tsujikawa, Properties of singularities in (phantom) dark energy universe, Phys. Rev. **D71**, 063004 (2005), arXiv:hep-th/0501025 [hep-th].

[35] T. Chiba, T. Okabe, and M. Yamaguchi, Kinetically driven quintessence, Phys. Rev. **D62**, 023511 (2000), arXiv:astro-ph/9912463 [astro-ph].

[36] C. Armendariz-Picon, V. F. Mukhanov, and P. J. Steinhardt, Essentials of k essence, Phys. Rev. **D63**, 103510 (2001), arXiv:astro-ph/0006373 [astro-ph].

[37] R.-G. Cai, A Dark Energy Model Characterized by the Age of the Universe, Phys. Lett. **B657**, 228 (2007), arXiv:0707.4049 [hep-th].

[38] H. Wei and R.-G. Cai, A New Model of Agegraphic Dark Energy, Phys. Lett. **B660**, 113 (2008), arXiv:0708.0884 [astro-ph].

[39] H. Wei and R.-G. Cai, Interacting Agegraphic Dark Energy, Eur. Phys. J. **C59**, 99 (2009), arXiv:0707.4052 [hep-th].

[40] A. G. Cohen, D. B. Kaplan, and A. E. Nelson, Effective field theory, black holes, and the cosmological constant, Phys. Rev. Lett. **82**, 4971 (1999), arXiv:hep-th/9803132 [hep-th].

[41] S. D. Thomas, Holography stabilizes the vacuum energy, Phys. Rev. Lett. **89**, 081301 (2002).

[42] M. Li, A Model of holographic dark energy, Phys. Lett. **B603**, 1 (2004), arXiv:hep-th/0403127 [hep-th].

[43] M. Novello and N. P. Neto, A Modified Theory Of Gravirt, 1987, CBPF-NF-002/87,, (1987).

[44] T. Clifton, P. G. Ferreira, A. Padilla, and C. Skordis, Modified Gravity and Cosmology, Phys. Rept. **513**, 1 (2012), arXiv:1106.2476 [astro-ph.CO].

[45] F. Pace, J. C. Waizmann, and M. Bartelmann, Spherical collapse model in dark energy cosmologies, Mon. Not. Roy. Astron. Soc. **406**, 1865 (2010), arXiv:1005.0233 [astro-ph.CO].

[46] R. R. Caldwell and E. V. Linder, Limits of quintessence, Phys. Rev. Lett. **95**, 141301 (2005).

[47] L. R. Abramo, R. C. Batista, L. Liberato, and R. Rosenfeld, Structure formation in the presence of dark energy perturbations, JCAP **0711**, 012, arXiv:0707.2882 [astro-ph].

[48] L. R. Abramo, R. C. Batista, L. Liberato, and R. Rosenfeld, Physical approximations for the nonlinear evolution of perturbations in inhomogeneous dark energy scenarios, Phys. Rev. **D79**, 023516 (2009), arXiv:0806.3461 [astro-ph].

[49] R. C. Batista and F. Pace, Structure formation in inhomogeneous Early Dark Energy models, JCAP **1306**, 044, arXiv:1303.0414 [astro-ph.CO].

[50] R. C. Batista, Impact of dark energy perturbations on the growth index, Phys. Rev. **D89**, 123508 (2014), arXiv:1403.2985 [astro-ph.CO].

[51] S. A. Appleby, E. V. Linder, and J. Weller, Cluster Probes of Dark Energy Clustering, Phys. Rev. **D88**, 043526 (2013), arXiv:1305.6982 [astro-ph.CO].

[52] F. Pace, R. C. Batista, and A. Del Popolo, Effects of shear and rotation on the spherical collapse model for clustering dark energy, Mon. Not. Roy. Astron. Soc. **445**, 648 (2014), arXiv:1406.1448 [astro-ph.CO].

[53] F. Pace, L. Moscardini, R. Crittenden, M. Bartelmann, and V. Pettorino, A comparison of structure formation in minimally and non-minimally coupled quintessence models, Mon. Not. Roy. Astron. Soc. **437**, 547 (2014), arXiv:1307.7026 [astro-ph.CO].

[54] T. M. Davis, E. Mörtsell, J. Sollerman, A. C. Becker, S. Blondin, P. Challis, A. Clocchiatti, A. Filippenko, R. Foley, P. M. Garnavich, *et al.*, Scrutinizing exotic cosmological models using essence supernova data combined with other cosmological probes, The Astrophysical Journal **666**, 716 (2007).

[55] W. M. Wood-Vasey, G. Miknaitis, C. Stubbs, S. Jha, A. Riess, P. M. Garnavich, R. P. Kirshner, C. Aguilera, A. C. Becker, J. Blackman, *et al.*, Observational

constraints on the nature of dark energy: first cosmological results from the essence supernova survey, The Astrophysical Journal **666**, 694 (2007).

[56] R. J. Scherrer and A. A. Sen, Thawing quintessence with a nearly flat potential, Phys. Rev. D **77**, 083515 (2008).

[57] R. J. Scherrer and A. Sen, Phantom dark energy models with a nearly flat potential, Physical Review D **78**, 067303 (2008).

[58] A. Ali, M. Sami, and A. Sen, Transient and late time attractor tachyon dark energy: Can we distinguish it from quintessence?, Physical Review D **79**, 123501 (2009).

[59] S. Sen, A. Sen, and M. Sami, The thawing dark energy dynamics: Can we detect it?, Physics Letters B **686**, 1 (2010).

[60] N. C. Devi and A. A. Sen, Evolution of spherical overdensity in thawing dark energy models, Monthly Notices of the Royal Astronomical Society **413**, 2371 (2011).

[61] D. F. Mota and C. van de Bruck, On the Spherical collapse model in dark energy cosmologies, Astron. Astrophys. **421**, 71 (2004), arXiv:astro-ph/0401504 [astro-ph].

[62] F. Felegary, I. Akhlaghi, and H. Haghi, Evolution of matter perturbations and observational constraints on tachyon scalar field model, Physics of the Dark Universe **30**, 100739 (2020).

[63] W. Yang, M. Shahalam, B. Pal, S. Pan, and A. Wang, Constraints on quintessence scalar field models using cosmological observations, Phys. Rev. D **100**, 023522 (2019).

[64] S. Sen, A. A. Sen, and M. Sami, The thawing dark energy dynamics: Can we detect it?, Phys. Lett. **B686**, 1 (2010), arXiv:0907.2814 [astro-ph.CO].

[65] E. J. Copeland, M. Sami, and S. Tsujikawa, Dynamics of dark energy, International Journal of Modern Physics D **15**, 1753 (2006).

[66] R. J. Scherrer and A. Sen, Thawing quintessence with a nearly flat potential, Physical Review D **77**, 083515 (2008).

[67] R. J. Scherrer and A. A. Sen, Phantom Dark Energy Models with a Nearly Flat Potential, Phys. Rev. **D78**, 067303 (2008), arXiv:0808.1880 [astro-ph].

[68] E. Gaztanaga, A. Cabre, and L. Hui, Clustering of Luminous Red Galaxies IV: Baryon Acoustic Peak in the Line-of-Sight Direction and a Direct Measurement of H(z), Mon. Not. Roy. Astron. Soc. **399**, 1663 (2009), arXiv:0807.3551 [astro-ph].

[69] M. Moresco *et al.*, Improved constraints on the expansion rate of the Universe up to z 1.1 from the spectroscopic evolution of cosmic chronometers, JCAP **1208**, 006, arXiv:1201.3609 [astro-ph.CO].

[70] C. Blake *et al.*, The WiggleZ Dark Energy Survey: Joint measurements of the expansion and growth history at z < 1, Mon. Not. Roy. Astron. Soc. **425**, 405 (2012), arXiv:1204.3674 [astro-ph.CO].

[71] L. Anderson *et al.* (BOSS), The clustering of galaxies in the SDSS-III Baryon Oscillation Spectroscopic Survey: baryon acoustic oscillations in the Data Releases 10 and 11 Galaxy samples, Mon. Not. Roy. Astron. Soc. **441**, 24 (2014), arXiv:1312.4877 [astro-ph.CO].

[72] C. Zhang, H. Zhang, S. Yuan, T.-J. Zhang, and Y.-C. Sun, Four new observational $H(z)$ data from luminous red galaxies in the Sloan Digital Sky Survey data release seven, Res. Astron. Astrophys. **14**, 1221 (2014), arXiv:1207.4541 [astro-ph.CO].

[73] R. Jimenez, L. Verde, T. Treu, and D. Stern, Constraints on the equation of state of dark energy and the Hubble constant from stellar ages and the CMB, Astrophys. J. **593**, 622 (2003), arXiv:astro-ph/0302560 [astro-ph].

[74] J. Simon, L. Verde, and R. Jimenez, Constraints on the redshift dependence of the dark energy potential, Phys. Rev. **D71**, 123001 (2005), arXiv:astro-ph/0412269 [astro-ph].

[75] M. Moresco, L. Pozzetti, A. Cimatti, R. Jimenez, C. Maraston, L. Verde, D. Thomas, A. Citro, R. Tojeiro, and D. Wilkinson, A 6% measurement of the Hubble parameter at $z \sim 0.45$: direct evidence of the epoch of cosmic re-acceleration, JCAP **1605** (05), 014, arXiv:1601.01701 [astro-ph.CO].

[76] M. Moresco, Raising the bar: new constraints on the Hubble parameter with cosmic chronometers at z 2, Mon. Not. Roy. Astron. Soc. **450**, L16 (2015), arXiv:1503.01116 [astro-ph.CO].

[77] D. Stern, R. Jimenez, L. Verde, M. Kamionkowski, and S. A. Stanford, Cosmic Chronometers: Constraining the Equation of State of Dark Energy. I: H(z) Measurements, JCAP **1002**, 008, arXiv:0907.3149 [astro-ph.CO].

[78] H. Yu, B. Ratra, and F.-Y. Wang, **856**, 3 (2018).

[79] D. M. Scolnic *et al.*, The Complete Light-curve Sample of Spectroscopically Confirmed SNe Ia from Pan-STARRS1 and Cosmological Constraints from the Combined Pantheon Sample, Astrophys. J. **859**, 101 (2018), arXiv:1710.00845 [astro-ph.CO].

[80] M. Sabiee, M. Malekjani, and D. Mohammad Zadeh Jassur, f(T) cosmology against the cosmographic method: A new study using mock and observational data, Mon. Not. Roy. Astron. Soc. **516**, 2597 (2022), arXiv:2212.04113 [astro-ph.CO].

[81] J. Garriga and V. F. Mukhanov, Perturbations in k-inflation, Physics Letters B **458**, 219 (1999).

[82] C. Armendariz-Picon, T. Damour, and V.-i. Mukhanov, k-inflation, Physics Letters B **458**, 209 (1999).

[83] F. Pace, L. Moscardini, R. Crittenden, M. Bartelmann, and V. Pettorino, A comparison of structure formation in minimally and non-minimally coupled quintessence models, Monthly Notices of the Royal Astronomical Society **437**, 547 (2014).

[84] C. Armendariz-Picon, V. Mukhanov, and P. J. Steinhardt, Dynamical solution to the problem of a small cosmological constant and late-time cosmic acceleration, Physical Review Letters **85**, 4438 (2000).

[85] J. K. Erickson, R. Caldwell, P. J. Steinhardt, C. Armendariz-Picon, and V. Mukhanov, Measuring the speed of sound of quintessence, Physical review letters **88**, 121301 (2002).

[86] R. Bean and O. Dore, Probing dark energy perturbations: the dark energy equation of state and speed of sound as measured by wmap, Physical Review D **69**, 083503 (2004).

[87] W. Hu and R. Scranton, Measuring dark energy clustering with cmb-galaxy correlations, Physical Review D **70**, 123002 (2004).

[88] G. Ballesteros and A. Riotto, Parameterizing the effect of dark energy perturbations on the growth of structures, Physics Letters B **668**, 171 (2008).

[89] S. Basilakos, J. C. B. Sanchez, and L. Perivolaropoulos, Spherical collapse model and cluster formation beyond the $\lambda$ cosmology: Indications for a clustered dark energy?, Physical Review D **80**, 043530 (2009).

[90] R. de Putter, D. Huterer, and E. V. Linder, Measuring the speed of dark: Detecting dark energy perturbations, Physical Review D **81**, 103513 (2010).

[91] R. Akhoury, D. Garfinkle, and R. Saotome, Gravitational collapse of k-essence, Journal of High Energy Physics **2011**, 96 (2011).

[92] D. Sapone and E. Majerotto, Fingerprinting dark energy. iii. distinctive marks of viscosity, Physical Review D **85**, 123529 (2012).

[93] A. Mehrabi, S. Basilakos, M. Malekjani, and Z. Davari, Growth of matter perturbations in clustered holographic dark energy cosmologies, Physical Review D **92**, 123513 (2015).

[94] L. R. Abramo, R. C. Batista, L. Liberato, and R. Rosenfeld, Structure formation in the presence of dark energy perturbations, Journal of Cosmology and Astroparticle Physics **2007** (11), 012.

[95] L. Abramo, R. Batista, L. Liberato, and R. Rosenfeld, Physical approximations for the nonlinear evolution of perturbations in inhomogeneous dark energy scenarios, Physical Review D **79**, 023516 (2009).

[96] R. Batista and F. Pace, Structure formation in inhomogeneous early dark energy models, Journal of Cosmology and Astroparticle Physics **2013** (06), 044.

[97] R. C. Batista, Impact of dark energy perturbations on the growth index, Physical Review D **89**, 123508 (2014).

[98] F. Pace, C. Fedeli, L. Moscardini, and M. Bartelmann, Structure formation in cosmologies with oscillating dark energy, Monthly Notices of the Royal Astronomical Society **422**, 1186 (2012).

[99] M. Setare, F. Felegary, and F. Darabi, Evolution of spherical over-densities in tachyon scalar field model, Physics Letters B **772**, 70 (2017).

[100] L. Abramo, R. Batista, L. Liberato, and R. Rosenfeld, Dynamical mutation of dark energy, Physical Review D **77**, 067301 (2008).

[101] F. Pace, J.-C. Waizmann, and M. Bartelmann, Spherical collapse model in dark-energy cosmologies, Monthly Notices of the Royal Astronomical Society **406**, 1865 (2010).

[102] A. Mehrabi, M. Malekjani, and F. Pace, Can observational growth rate data favor the clustering dark energy models?, Astrophys. Space Sci. **356**, 129 (2015), arXiv:1411.0780 [astro-ph.CO].

[103] A. Mehrabi, S. Basilakos, and F. Pace, How clustering dark energy affects matter perturbations, Mon. Not. Roy. Astron. Soc. **452**, 2930 (2015), arXiv:1504.01262 [astro-ph.CO].

[104] M. Malekjani, S. Basilakos, Z. Davari, A. Mehrabi, and M. Rezaei, Agegraphic dark energy: growth index and cosmological implications, Mon. Not. Roy. Astron. Soc. **464**, 1192 (2017), arXiv:1609.01998 [astro-ph.CO].

[105] V. Silveira and I. Waga, Decaying $\Lambda$ cosmologies and power spectrum, Phys. Rev. D **50**, 4890 (1994).

[106] M. Rezaei, M. Malekjani, S. Basilakos, A. Mehrabi, and D. F. Mota, Constraints to Dark Energy Using PADE Parameterizations, Astrophys. J. **843**, 65 (2017), arXiv:1706.02537 [astro-ph.CO].

[107] On the growth of linear perturbations, Physics Letters B **660**, 439 (2008).

[108] D. Huterer, D. Shafer, D. Scolnic, and F. Schmidt, Testing $\Lambda$CDM at the lowest redshifts with SN Ia and galaxy velocities, JCAP **1705** (05), 015, arXiv:1611.09862 [astro-ph.CO].

[109] M. J. Hudson and S. J. Turnbull, The growth rate of cosmic structure from peculiar velocities at low and high redshifts, Astrophys. J. **751**, L30 (2013), arXiv:1203.4814 [astro-ph.CO].

[110] S. J. Turnbull, M. J. Hudson, H. A. Feldman, M. Hicken, R. P. Kirshner, and R. Watkins, Cosmic flows in the nearby universe from Type Ia Supernovae, Mon. Not. Roy. Astron. Soc. **420**, 447 (2012), arXiv:1111.0631 [astro-ph.CO].

[111] M. Davis, A. Nusser, K. Masters, C. Springob, J. P. Huchra, and G. Lemson, Local Gravity versus Local Velocity: Solutions for $\beta$ and nonlinear bias, Mon. Not. Roy. Astron. Soc. **413**, 2906 (2011), arXiv:1011.3114 [astro-ph.CO].

[112] M. Feix, A. Nusser, and E. Branchini, Growth Rate of Cosmological Perturbations at z0.1 from a New Observational Test, Phys. Rev. Lett. **115**, 011301 (2015), arXiv:1503.05945 [astro-ph.CO].

[113] C. Howlett, A. Ross, L. Samushia, W. Percival, and M. Manera, The clustering of the SDSS main galaxy sample II. Mock galaxy catalogues and a measurement of the growth of structure from redshift space distortions at $z = 0.15$, Mon. Not. Roy. Astron. Soc. **449**, 848 (2015), arXiv:1409.3238 [astro-ph.CO].

[114] Y.-S. Song and W. J. Percival, Reconstructing the history of structure formation using Redshift Distortions, JCAP **0910**, 004, arXiv:0807.0810 [astro-ph].

[115] C. Blake *et al.*, Galaxy And Mass Assembly (GAMA): improved cosmic growth measurements using multiple tracers of large-scale structure, Mon. Not. Roy. Astron. Soc. **436**, 3089 (2013), arXiv:1309.5556 [astro-ph.CO].

- [116] L. Samushia, W. J. Percival, and A. Raccanelli, Interpreting large-scale redshift-space distortion measurements, Mon. Not. Roy. Astron. Soc. **420**, 2102 (2012), arXiv:1102.1014 [astro-ph.CO].
- [117] A. G. Sanchez *et al.*, The clustering of galaxies in the SDSS-III Baryon Oscillation Spectroscopic Survey: cosmological implications of the full shape of the clustering wedges in the data release 10 and 11 galaxy samples, Mon. Not. Roy. Astron. Soc. **440**, 2692 (2014), arXiv:1312.4854 [astro-ph.CO].
- [118] C.-H. Chuang *et al.*, The clustering of galaxies in the SDSS-III Baryon Oscillation Spectroscopic Survey: single-probe measurements from CMASS anisotropic galaxy clustering, Mon. Not. Roy. Astron. Soc. **461**, 3781 (2016), arXiv:1312.4889 [astro-ph.CO].
- [119] A. Pezzotta *et al.*, The VIMOS Public Extragalactic Redshift Survey (VIPERS): The growth of structure at $0.5 < z < 1.2$ from redshift-space distortions in the clustering of the PDR-2 final sample, Astron. Astrophys. **604**, A33 (2017), arXiv:1612.05645 [astro-ph.CO].
- [120] T. Okumura *et al.*, The Subaru FMOS galaxy redshift survey (FastSound). IV. New constraint on gravity theory from redshift space distortions at $z \sim 1.4$, Publ. Astron. Soc. Jap. **68**, 38 (2016), arXiv:1511.08083 [astro-ph.CO].
- [121] P. Bessa, M. Campista, and A. Bernui, Observational constraints on starobinsky f(r) cosmology from cosmic expansion and structure growth data, The European Physical Journal C **82**, 10.1140/epjc/s10052-022-10457-z (2022).
- [122] G. Hinshaw, D. Larson, E. Komatsu, D. N. Spergel, C. L. Bennett, J. Dunkley, M. R. Nolta, M. Halpern, R. S. Hill, N. Odegard, L. Page, K. M. Smith, J. L. Weiland, B. Gold, N. Jarosik, A. Kogut, M. Limon, S. S. Meyer, G. S. Tucker, E. Wollack, and E. L. Wright, NINE-YEAR iWILKINSON MICROWAVE ANISOTROPY PROBE/i ( iWMAP/i ) OBSERVATIONS: COSMOLOGICAL PARAMETER RESULTS, The Astrophysical Journal Supplement Series **208**, 19 (2013).
- [123] F. Beutler, C. Blake, M. Colless, D. H. Jones, L. Staveley-Smith, L. Campbell, Q. Parker, W. Saunders, and F. Watson, The 6dF Galaxy Survey: Baryon Acoustic Oscillations and the Local Hubble Constant, Mon. Not. Roy. Astron. Soc. **416**, 3017 (2011), arXiv:1106.3366 [astro-ph.CO].
- [124] N. Padmanabhan, X. Xu, D. J. Eisenstein, R. Scalzo, A. J. Cuesta, K. T. Mehta, and E. Kazin, A 2 per cent distance to $z$=0.35 by reconstructing baryon acoustic oscillations - I. Methods and application to the Sloan Digital Sky Survey, Mon. Not. Roy. Astron. Soc. **427**, 2132 (2012), arXiv:1202.0090 [astro-ph.CO].
- [125] L. Anderson *et al.*, The clustering of galaxies in the SDSS-III Baryon Oscillation Spectroscopic Survey: Baryon Acoustic Oscillations in the Data Release 9 Spectroscopic Galaxy Sample, Mon. Not. Roy. Astron. Soc. **427**, 3435 (2013), arXiv:1203.6594 [astro-ph.CO].
- [126] G. Hinshaw *et al.* (WMAP), Nine-Year Wilkinson Microwave Anisotropy Probe (WMAP) Observations: Cosmological Parameter Results, Astrophys. J. Suppl. **208**, 19 (2013), arXiv:1212.5226 [astro-ph.CO].
- [127] D. L. Shafer and D. Huterer, Chasing the phantom: A closer look at Type Ia supernovae and the dark energy equation of state, Phys. Rev. **D89**, 063510 (2014), arXiv:1312.1688 [astro-ph.CO].
- [128] S. Burles, K. M. Nollett, and M. S. Turner, Big bang nucleosynthesis predictions for precision cosmology, Astrophys. J. **552**, L1 (2001), arXiv:astro-ph/0010171 [astro-ph].
- [129] P. Serra, A. Cooray, D. E. Holz, A. Melchiorri, S. Pandolfi, and D. Sarkar, No Evidence for Dark Energy Dynamics from a Global Analysis of Cosmological Data, Phys. Rev. **D80**, 121302 (2009), arXiv:0908.3186 [astro-ph.CO].
- [130] S. Basilakos, M. Plionis, and J. Solà, Hubble expansion & Structure Formation in Time Varying Vacuum Models, Phys. Rev. **D80**, 083511 (2009), arXiv:0907.4555 [astro-ph.CO].
- [131] A. Mehrabi, S. Basilakos, M. Malekjani, and Z. Davari, Growth of matter perturbations in clustered holographic dark energy cosmologies, Phys. Rev. **D92**, 123513 (2015), arXiv:1510.03996 [astro-ph.CO].
- [132] H. Akaike, A new look at the statistical model identification, IEEE transactions on automatic control **19**, 716 (1974).
- [133] G. Schwarz, Estimating the dimension of a model, The annals of statistics **6**, 461 (1978).
- [134] A. Cooray, D. Huterer, and D. Baumann, Growth rate of large scale structure as a powerful probe of dark energy, Phys. Rev. **D69**, 027301 (2004), arXiv:astro-ph/0304268 [astro-ph].
- [135] P.-S. Corasaniti, T. Giannantonio, and A. Melchiorri, Constraining dark energy with cross-correlated CMB and large scale structure data, Phys. Rev. **D71**, 123521 (2005), arXiv:astro-ph/0504115 [astro-ph].
- [136] T. Koivisto and D. F. Mota, Cosmology and Astrophysical Constraints of Gauss-Bonnet Dark Energy, Phys. Lett. **B644**, 104 (2007), arXiv:astro-ph/0606078 [astro-ph].
- [137] D. F. Mota, J. R. Kristiansen, T. Koivisto, and N. E. Groeneboom, Constraining Dark Energy Anisotropic Stress, Mon. Not. Roy. Astron. Soc. **382**, 793 (2007), arXiv:0708.0830 [astro-ph].
- [138] D. F. Mota, D. J. Shaw, and J. Silk, On the Magnitude of Dark Energy Voids and Overdensities, Astrophys. J. **675**, 29 (2008), arXiv:0709.2227 [astro-ph].
- [139] S. Basilakos, M. Plionis, and J. A. S. Lima, Confronting Dark Energy Models using Galaxy Cluster Number Counts, Phys. Rev. **D82**, 083517 (2010), arXiv:1006.3418 [astro-ph.CO].
- [140] R. Gannouji, B. Moraes, D. F. Mota, D. Polarski, S. Tsujikawa, and H. A. Winther, Chameleon dark energy models with characteristic signatures, Phys.

Rev. **D82**, 124006 (2010), arXiv:1010.3769 [astro-ph.CO].

[141] S. Nesseris, C. Blake, T. Davis, and D. Parkinson, The WiggleZ Dark Energy Survey: constraining the evolution of Newton's constant using the growth rate of structure, JCAP **1107**, 037, arXiv:1107.3659 [astro-ph.CO].

[142] F. Pace, C. Fedeli, L. Moscardini, and M. Bartelmann, Structure formation in cosmologies with oscillating dark energy, Mon. Not. Roy. Astron. Soc. **422**, 1186 (2012), arXiv:1111.1556 [astro-ph.CO].

[143] S. Basilakos and A. Pouri, The growth index of matter perturbations and modified gravity, Mon. Not. Roy. Astron. Soc. **423**, 3761 (2012), arXiv:1203.6724 [astro-ph.CO].

[144] C.-H. Chuang *et al.*, The clustering of galaxies in the SDSS-III Baryon Oscillation Spectroscopic Survey: single-probe measurements and the strong power of normalized growth rate on constraining dark energy, Mon. Not. Roy. Astron. Soc. **433**, 3559 (2013), arXiv:1303.4486 [astro-ph.CO].

[145] C. Llinares and D. Mota, Releasing scalar fields: cosmological simulations of scalar-tensor theories for gravity beyond the static approximation, Phys. Rev. Lett. **110**, 161101 (2013), arXiv:1302.1774 [astro-ph.CO].

[146] W. Yang, L. Xu, Y. Wang, and Y. Wu, Constraints on a decomposed dark fluid with constant adiabatic sound speed by jointing the geometry test and growth rate after Planck data, Phys. Rev. **D89**, 043511 (2014), arXiv:1312.2769 [astro-ph.CO].

[147] C. Llinares, D. F. Mota, and H. A. Winther, ISIS: a new N-body cosmological code with scalar fields based on RAMSES. Code presentation and application to the shapes of clusters, Astron. Astrophys. **562**, A78 (2014), arXiv:1307.6748 [astro-ph.CO].

[148] J. Li, R. Yang, and B. Chen, Discriminating dark energy models by using the statefinder hierarchy and the growth rate of matter perturbations, JCAP **1412** (12), 043, arXiv:1406.7514 [gr-qc].

[149] I. A. Akhlaghi, M. Malekjani, S. Basilakos, and H. Haghi, Model selection and constraints from Holographic dark energy scenarios, Mon. Not. Roy. Astron. Soc. **477**, 3659 (2018), arXiv:1804.02989 [gr-qc].

[150] L. Chen, Q.-G. Huang, and K. Wang, Distance Priors from Planck Final Release, JCAP **02**, 028, arXiv:1808.05724 [astro-ph.CO].